\documentclass[12pt]{article}

\usepackage{latexsym}

\begin{document}

\title{Low-frequency
scalar absorption cross sections for stationary black holes}

\author{Atsushi Higuchi\\
{\normalsize Department of Mathematics, University of York}\\ 
{\normalsize Heslington, York, YO10 5DD, UK}\\
{\normalsize Email: ah28@york.ac.uk}}

\date{1 November, 2001}

\maketitle

\begin{abstract}
We discuss the absorption cross section for the minimally-coupled
massless scalar field into a stationary and circularly symmetric black hole 
with nonzero angular velocity
in four or higher dimensions.  In particular, we show that it
equals the horizon area
in the zero-frequency limit provided that 
the solution of the scalar field equation
with an incident monochromatic plane wave converges pointwise
to a smooth time-independent solution
outside the black hole and on the future horizon,
with the error term being at most linear in the frequency.
We also show that
this equality holds for static black holes which are not necessarily 
spherically symmetric. 
The zero-frequency scattering cross section is found to vanish in both cases.
It is shown in an Addendum that the equality holds for any stationary black
hole with vanishing expansion if the limit solution is known to be a constant.
\end{abstract}

Properties of black holes have been attracting much attention
in connection with string theory since it was found that the Bekenstein-Hawking
entropy is reproduced by string theory for some black holes~\cite{SV}.
The absorption cross section for the minimally-coupled massless scalar field
was studied in this context~\cite{DM1,DM2,GK,MS1}, and 
Das, Gibbons and Mathur showed that it is exactly the horizon area for
a static and spherically symmetric black hole in any dimensions 
in the zero-frequency
limit~\cite{DGM}. (See, e.g.~\cite{Staro,Gib,Pag,Unr} for earlier calculations
of absorption cross sections for black holes.) 
It is natural to ask whether the same result holds for rotating black holes.
In fact Maldacena and Strominger have shown~\cite{MS}, among other things,
that this equality 
is valid for the four-dimensional Kerr-Newman black hole. 
This result has been generalized, and the equality has been
verified for 
other rotating black holes as well~\cite{CL1,CL2,GH}. 
However, there has been no general explanation of
this universality so far except for the original work
dealing with the static and spherically symmetric case. 
In this Letter we show that this equality holds
for any stationary and circularly symmetric black hole
with nonzero angular velocity
in any spacetime dimensions larger than three
{\it if the solution of the scalar field equation
with an incident monochromatic plane wave 
converges pointwise to a smooth
time-independent solution in the zero-frequency
limit outside the black hole and on the future horizon, with the error 
term being at most linear in the frequency}.
[We call the solution with an incident monochromatic plane wave
the asymptotically-plane-wave (APW) solution.]
We also show the same equality for any static black hole, which may not
be spherically symmetric.  Then we show that the
zero-frequency scattering cross section vanishes for these black holes.

By definition a static black-hole spacetime admits a Killing vector field
(generating time translations) 
which is hypersurface-orthogonal, timelike outside
the black hole (or in the domain of outer communications~\cite{Carter}) and
null on the future and past horizons. 
Thus, the metric 
can be written outside the black hole as 
\begin{equation}
ds^2|_{\rm static} = -N^2 dt^2 + h^{(s)}_{mn}dx^m dx^n\,,
\label{static}
\end{equation}
where $N^2\,(>0)$ and $h^{(s)}_{mn}$ (which is a positive-definite matrix) are 
independent of $t$. 
In $(p+2)$-dimensional stationary and circularly symmetric black-hole
spacetime, the $2$-surfaces generated by the commuting 
Killing vectors $(\partial/\partial t)^a$
and $(\partial/\partial\phi)^a$ are orthogonal to another
family of $p$-dimensional surfaces by definition~\cite{Carter,Carter2}. 
({\it We assume that the latter surfaces are spacelike outside the black hole
and on the horizons.})
Moreover, the
spacetime is periodic in $\phi$ with period $2\pi$.  Thus, the metric
can be made 
$t$- and $\phi$-independent and written outside the black hole as 
\begin{equation}
ds^2|_{\rm circular} = -N^2 dt^2 + h_{\phi\phi}(d\phi - N^\phi dt)^2
+ h^{(c)}_{mn}dx^m dx^n\,, \label{circular}
\end{equation}
where the function $h_{\phi\phi}$ and the matrix $h^{(c)}_{mn}$ are
positive definite.  

{\it In both cases, we assume that
compact spacelike $p$-surfaces of constant $t$ foliate the future 
horizon\footnote{The variable $t$
here differs from that in (\ref{static}) or (\ref{circular}) by
addition of a function of the other variables.}
and that the spacetime is asymptotically flat.
In the circularly symmetric case,
we assume that the null generator of the future horizon is of the form
$(\partial/\partial t)^a + \Omega_H(\partial/\partial \phi)^a$ 
with the constant $\Omega_H$ being nonzero,
and that the $p$-surfaces foliating the future horizon are invariant under
$\phi$-rotations.}\footnote{We will not explicitly 
use the corresponding statements for the past horizon.}
[A stationary black hole
in general relativity in four dimensions is either static or circularly 
symmetric and
satisfies the conditions required here
under reasonable assumptions (see \cite{Hawking,Carter}).]

Let us first derive a useful formula for the absorption cross section
for any stationary black hole.  
Far away from the black hole, the APW solution,
$\psi = \Psi_\omega e^{-i\omega t}$, $\omega > 0$, of the equation
$\nabla_a \nabla^a \psi = 0$
in $p+2$ dimensions is approximately given by
\begin{equation}
\Psi_\omega \approx e^{i\omega r\cos\theta} 
+ f_\omega(\Omega)\,\frac{e^{i\omega r}}{r^{p/2}}\,.
\label{far}
\end{equation}
Here, $\Omega$ represents all angular variables on the unit $p$-sphere, and
$r$ is the approximate
distance from the black hole.  The variable $\theta$ is the
angle between 
the position vector and the direction in which the incident wave propagates. 
There is a conserved current given by
\begin{equation}
J_a = -i \left( \psi^*\nabla_a \psi
- \psi \nabla_a \psi^*\right)\,, \label{jmu}
\end{equation}
where the star denotes complex conjugation.  {\it The current flowing out
of the {\it past}
horizon must be zero for the APW solution.}
This requirement will be important later.
For the incident wave
$e^{i\omega r\cos\theta}$ in (\ref{far})
we have $n^a J_a = - J_t = 2\omega$ with $n^a$ being the unit vector in the
direction of propagation.  
Therefore, the absorption cross section is $(2\omega)^{-1}$ times
the rate of this current going across the future horizon integrated over the
horizon area.

A stationary black hole has the future and past horizons 
which are null and fixed under time translations. 
The spacetime metric in a neighbourhood of 
the future horizon, parametrized by $x^i$ $(i=1,\ldots,p)$,
$\lambda$ and $t$, can be chosen as
\begin{equation}
ds^2 = -Adt^2 + g_{ij}(dx^i - C^i dt)(dx^j - C^jdt) + 2Bd\lambda dt\,,
\label{metric1}
\end{equation}
where $A=0$ at $\lambda=0$ (on the horizon) and 
$A>0$ for $\lambda>0$ (outside the
horizon).  The quantities
$B\,(>0)$, $A$, $C^i$ and the positive-definite matrix 
$g_{ij}$ are $t$-independent. 
This metric can be obtained as follows.  Let the line element of the compact 
spacelike 
$p$-surfaces, $S_t$, of constant $t$, which foliate the future horizon, be
$g_{ij}dx^i dx^j$.  Then the line element 
on the horizon is $g_{ij}(dx^i - C^i dt)(dx^j - C^j dt)$ for some $C^i$.
Consider the past-directed and outward-pointing null geodesics which 
are orthogonal to $S_t$.
Let $t$ and $x^i$ be constant on each of these geodesics
and let $\lambda$ be its affine parameter. Then the metric is of the form 
(\ref{metric1}) since these geodesics are orthogonal to
$(\partial/\partial x^i)^a$ for all $\lambda$.

A $(p+1)$-dimensional hypersurface of constant $\lambda$
is timelike for $\lambda >0$ and null for $\lambda =0$.
If we write the unit normal to the constant $\lambda$ hypersurface with 
$\lambda > 0$ which points towards the horizon as $-A^{-1/2} l^a$, then
\begin{equation}
l^a = 
(\partial/\partial t)^a
+ B^{-1}A(\partial/\partial\lambda)^a
+ C^i (\partial/\partial x^i)^a\,.
\end{equation} 
Note that the vector $l^a$ on the horizon,
\begin{equation}
l^a|_{\rm horizon} = 
(\partial/\partial t)^a
+ C^i(\partial/\partial x^i)^a\,,
\label{horla}
\end{equation}
is the null generator of the horizon.
The volume element of the hypersurface of constant $\lambda$ is 
$A^{1/2}\sqrt{g}$.  Therefore, the absorption cross section is
\begin{eqnarray}
\sigma_\omega & = & - \frac{1}{2\omega}\int d^p x \sqrt{g}\, 
l^a J_a \nonumber \\
& = & \int d^p x \sqrt{g} |\Psi_\omega|^2
- \frac{1}{\omega}\,{\rm Im}\int d^p x \sqrt{g}\,C^i 
\Psi_\omega^* \partial_i \Psi_\omega\,, \label{keyeq}
\end{eqnarray}
where integration is performed on the future horizon.
If
\begin{equation}
\lim_{\omega \to 0}|\Psi_\omega|^2 = 1 \label{problem1}
\end{equation}
and
\begin{equation}
\lim_{\omega \to 0}\frac{1}{\omega}\,\int d^p x\,\sqrt{g}\,C^i
\Psi^*_\omega \partial_i \Psi_\omega = 0 \label{problem2}
\end{equation}
on the future horizon,
then it follows from (\ref{keyeq}) that
the zero-frequency absorption cross section $\sigma_0$ equals the horizon area
$\int d^p x \sqrt{g}$. 
We will show that equations (\ref{problem1}) and
(\ref{problem2}) indeed hold under the assumptions stated before.
We will first assume in addition that 
\begin{equation}
\Psi_0 \equiv \lim_{\omega \to 0}\Psi_\omega \approx 1 + {\cal O}(r^{-p/2})
\label{falloff}
\end{equation}
for large $r$, where ${\cal O}(r^{-\beta})$ denotes a term of order
$r^{-\beta}$ or smaller.
This assumption will be justified later for $p\geq 2$.\footnote{
We have been unable
to establish (\ref{falloff}) for $p=1$, 
and this is why our result is limited to
four or higher dimensions.}

For the static case we have $C^i = 0$ on the horizon
because the Killing vector $(\partial/\partial t)^a$ 
is orthogonal to the horizon.
Thus, equation (\ref{problem2}) is trivially satisfied. 
Now, $f_0 \equiv \Psi_0 - 1$ is a time-independent solution 
of $\nabla_a \nabla^a f_0 = 0$ falling off at least as fast as $r^{-p/2}$
for large $r$ due to (\ref{falloff}).  
(We assume that $f_0$ is real.  If this is not the case, one can consider the
real and imaginary parts of $f_0$ separately.)
Consider the spacetime integral 
of $\nabla_a f_0 \nabla^a f_0$ {\it in the region outside the horizon}
between two spacelike hypersurfaces which intersect the future horizon and 
are orthogonal to
$(\partial/\partial t)^a$ far away from the black hole.
Let one of the hypersurfaces be obtained by a time
translation of the other.  Then  we have
\begin{eqnarray}
\int dV \nabla_a f_0 \nabla^a f_0 & = &
\int dV \nabla_a (f_0 \nabla^a f_0) \nonumber \\
& = &
-\int_{\rm horizon} dt\,d^p x\,\sqrt{g}
f_0 \frac{\partial f_0}{\partial t}
+ \lim_{r\to \infty}\int dt\,d\Omega\, 
r^p\,f_0 \frac{\partial f_0}{\partial r}\,, \label{beken1}
\end{eqnarray}
where $dV$ is the spacetime volume element.  The first term on the right-hand
side vanishes
because $f_0$ is time-independent.  The second term is zero because
of the fall-off condition (\ref{falloff}). 
However, $\nabla_a f_0 \nabla^a f_0$ is non-negative 
outside the black hole because
\[
\nabla_a f_0 \nabla^a f_0  = h^{(s)mn}\partial_m f_0 \partial_n f_0\,,
\]
where $h^{(s)mn}$ is the inverse of $h^{(s)}_{mn}$ in (\ref{static}).
Thus, equation (\ref{beken1}) implies that $\nabla_a f_0 = 0$.  Since
$f_0 \to 0$ as $r \to \infty$,  we must have $f_0 = 0$, i.e.
$\Psi_0 = 1$ outside the black hole.  (This is 
a special case of a no-hair theorem due to Bekenstein~\cite{BE1}.) 
Hence, equation (\ref{problem1}) holds on the future horizon 
by the assumed smoothness of the solution.

Next let us discuss the case with a stationary and circularly symmetric
black hole.
Here, we choose $\phi$ as one
of the $x^i$'s in (\ref{metric1}).  Then the metric $g_{ij}$ 
is $\phi$-invariant by assumption.
As we stated before, we assume that the null generator (\ref{horla}) of the
future horizon is 
$(\partial/\partial t)^a + \Omega_H(\partial/\partial \phi)^a$, where
$\Omega_H$ is a nonzero constant interpreted as the angular velocity of the
horizon.  Thus, we have on the horizon
\begin{equation}
C^i (\partial/\partial x^i)^a =
\Omega_H (\partial/\partial \phi)^a\,. \label{const}
\end{equation}
Let us write 
\[
\Psi_0 \equiv \sum_{m=-\infty}^{+\infty} \Psi_0^{(m)}\,,
\]
where $\partial_\phi \Psi_0^{(m)} = im\Psi_0^{(m)}$ with 
$\partial_\phi = \partial/\partial \phi$.
The (suitably normalized)
component of the current going into the future horizon for the solution
$\Psi_0^{(m)}$ is
$-l^a J_a = -2m\Omega_H|\Psi_0^{(m)}|^2$. 
By time independence of the solution we conclude that this equals
the current coming out of the past horizon.
Thus, for $m\neq 0$, the part proportional to $e^{im\phi}$ of the
current coming out of the past horizon would be nonzero
if $\Psi_0^{(m)}\neq 0$ on the future horizon.  
Since this could not be the case
for (the $\omega \to 0$ limit of) the APW solution,
we conclude that $\Psi_0^{(m)}= 0$ for all nonzero $m$.
Therefore, we have $\partial_\phi\Psi_0 = 0$ and
$\Psi_0 = \Psi_0^{(0)}$ on the future horizon.
Then, $\partial_\phi \Psi_\omega = {\cal O}(\omega)$ for small $\omega$ 
by the assumption that the error term is at most linear in $\omega$.  Thus,
equation (\ref{problem2}) is satisfied because 
\begin{equation}
\int d^{p}x\,\sqrt{g}\,C^i\Psi_\omega^* \partial_i \Psi_\omega
= \Omega_H
\int d^{p}x\,\sqrt{g}\,\Psi_\omega^*\partial_\phi \Psi_\omega \nonumber
 =  {\cal O}(\omega^2)
\end{equation}
due to (\ref{const}).
Now, outside the black hole we find from (\ref{circular}) 
\[
\nabla_a f_0 \nabla^a f_0
= h^{(c)mn}\partial_m f_0 \partial_n f_0
+ \left[h^{\phi\phi}-(N^\phi)^2/N^2\right](\partial_\phi f_0)^2\,,
\]
where $h^{(c)mn}$ is the inverse of $h^{(c)}_{mn}$ in (\ref{circular}) and 
$h^{\phi\phi} \equiv (h_{\phi\phi})^{-1}$.
Therefore, $\nabla_a f_0 \nabla^a f_0$ 
is non-negative because $\partial_\phi f_0 = 0$.
Hence, we must have $f_0=0$ and $\Psi_0 = 1$ outside the black hole 
because essentially the same argument as in the static case is valid~\cite{BE2}.
Thus, equation (\ref{problem1}) holds
on the future horizon
again by the assumed smoothness.

Next, let us show equation (\ref{falloff}), which enabled us to use 
the no-hair theorems, for $p\geq 2$.
Far away from the horizon, the solution can be expanded as
\begin{equation}
\Psi_\omega \approx \sum_{l=0}^\infty R_{l\omega}(r)Y_l(\Omega)\,,
\end{equation}
where $Y_l(\Omega)$ is one of the spherical harmonics satisfying 
$\tilde{\Delta}Y_{l}(\Omega) = -l(l+p-1)Y_{l}(\Omega)$.
Here, $\tilde{\Delta}$ is the Laplacian on the unit $p$-sphere.
The radial function $R_{l\omega}(r)$ can be written as 
\begin{eqnarray}
R_{l\omega}(r) & \approx & c_{lp} (\omega r)^{-(p-1)/2}
\left[ (1+ \alpha_{l\omega})J_{l+(p-1)/2}(\omega r) - i(1-\alpha_{l\omega})
N_{l+(p-1)/2}(\omega r)
\right] \nonumber \\
& = & 
c_{lp} (\omega r)^{-(p-1)/2}
\left[ H^{(1)*}_{l+(p-1)/2}(\omega r) + \alpha_{l\omega} 
H^{(1)}_{l+(p-1)/2}(\omega r)\right]\,. \label{Rl}
\end{eqnarray}
Here, the functions
$J_\nu(x)$ and $N_\nu(x)$ are the Bessel and Neumann functions,
respectively, and 
$H_{\nu}^{(1)}(x) = J_\nu (x) + iN_\nu(x)$
is the Hankel function of the first kind. (See, e.g. \cite{GR} for the
definitions of these functions.)
The constants $c_{lp}$, 
which depend only on $l$ and $p$, are determined from the 
expansion of the plane wave $e^{i\omega r\cos\theta}$, 
and the first term in the second 
line of (\ref{Rl}) is independent of the details of spacetime
because it represents the wave coming
in from $r = \infty$.

Now, the function $N_\nu(x)$ behaves like $x^{-\nu}$
whereas the function $J_\nu(x)$ behaves like $x^\nu$ for small $x$ 
if $\nu > 0$.  
Therefore, for $R_{l\omega}(r)$ given by (\ref{Rl}) to have
a finite limit as $\omega \to 0$ for large $r$, we must have 
$1-\alpha_{l\omega} = {\cal O}(\omega^{l+p-1})$.
Then, we find
\begin{equation}
\lim_{\omega \to 0}R_{l\omega}(r) \approx {\cal O}(r^{-l-p+1}) \label{nonzerol}
\end{equation}
 for large $r$ if $l > 0$.
To find the behaviour of the $s$-wave radial function,
$R_{0\omega}(r)$, we note first that 
$\sum_{l=0}^\infty R_{l\omega}(r)Y_l(\Omega) = e^{i\omega r\cos\theta}$ if
$\alpha_{l\omega} = 1$ for all $l$ and $\omega$.  This implies that
\begin{equation}
\lim_{r\to 0}2c_{0p} (\omega r)^{-(p-1)/2}J_{(p-1)/2}(\omega r)Y_0(\Omega)
= 1\,.
\end{equation}
Since the $r\to 0$ and $\omega \to 0$ limits are equivalent on the 
left-hand side of this equation and since 
$\alpha_{0\omega} \to 1$ as $\omega \to 0$, 
we find
\begin{equation}
\lim_{\omega\to 0}R_{0\omega}(r)Y_0(\Omega) \approx 
1 + {\cal O}(r^{-(p-1)}) \label{zerol}
\end{equation} 
for large $r$.
We deduce equation (\ref{falloff}) from (\ref{nonzerol}) and (\ref{zerol})
because $l+p-1 \geq p/2$ if $p\geq 2$.

Finally, let us show that the scattering cross section vanishes.  
The contribution of the $l$-th partial wave to the scattering cross
section is at most of order $\omega^{2l+p-2}$ for small $\omega$
because it is a constant multiple of
$\omega^{-p}|1-\alpha_{l\omega}|^2$. 
[The contribution to the absorption cross section is 
$\omega^{-p}(1-|\alpha_{l\omega}|^2)$ times a numerical constant.]
Hence, it vanishes in the zero-frequency limit unless $p=2$ and $l=0$.
For $p=2$, i.e. in four dimensions, we find
$\Psi_0 \approx 1 + K/r$
with the scattering cross section being $4\pi|K|^2$.  Since $K=0$, the 
$s$-wave contribution to the scattering cross section must also vanish.

\section*{Addendum}

In fact, if the APW solution is known to converge to one (after a suitable
adjustment of the phase), then the equality of the zero-frequency absorption
cross section and the horizon area holds for any stationary black hole with
vanishing expansion.
(It is well known that
stationary black holes must have horizons with vanishing 
expansion in general relativity if appropriate
energy conditions are satisfied~\cite{Hawking,Carter}.) 

Let us first show that the horizon area is well defined if and only if
$D_i C^i = 0$ on the horizon, where $D_i$ is the convariant derivative 
compatible with the metric $g_{ij}$.
Note that the metric (\ref{metric1}) will still be $t$-independent after an
(infinitesimal) transformation
$t \to t + \epsilon$,
where $\epsilon$ is $x^i$-dependent but $t$-independent.  The metric
$g_{ij}$ changes as
$g_{ij} \to g_{ij} - C_i D_j\epsilon - C_j D_i\epsilon$ under this 
transformation.
Then the horizon area changes as
\begin{equation}
\int d^p x\,\sqrt{g} \to \int d^p x\,\sqrt{g}\,(1 + \epsilon D_i C^i)\,.
\end{equation}
Hence, the horizon area is well defined if and only if $D_i C^i = 0$.  
One can show that 
the expansion of the horizon generators is
proportional to $D_i C^i$. Therefore, the horizon area is well defined
if and only if the expansion of the horizon generators 
vanishes.  
Now, if $D_i C^i = 0$, then equation (\ref{problem2}) holds
because 
\begin{eqnarray}
\int d^p x\,\sqrt{g}\, C^i \Psi_\omega^*D_i \Psi_\omega
& = & 
\int d^p x\,\sqrt{g}\,C^i \Psi_\omega^*D_i (\Psi_\omega - 1) \nonumber \\
& = & 
- \int d^p x\,\sqrt{g}\,C^i(\Psi_\omega - 1)D_i\Psi_\omega^*
=  {\cal O}(\omega^2)\,.
\end{eqnarray}
Hence,
\begin{equation}
\lim_{\omega \to 0} \sigma_\omega
= \int d^p x \sqrt{g}|\Psi_\omega|^2
= \int d^p x \sqrt{g}\,.
\end{equation}

\

\begin{flushleft}
{\large {\bf Acknowledgements}}
\end{flushleft}

The author is grateful to John Friedman whose suggestions were essential
in initiating this work.  He also thanks Chris Fewster, Gary Gibbons
and Bernard Kay for useful discussions.

\end{document}